\documentclass[nohyperref]{article}
\usepackage{microtype}
\usepackage{graphicx}
\usepackage{subfigure}
\usepackage{booktabs} 

\usepackage{hyperref}

\usepackage{amsmath}
\usepackage{amssymb}
\usepackage{mathtools}
\usepackage{amsthm}

\usepackage[capitalize,noabbrev]{cleveref}

\usepackage[accepted]{icml2022}

\usepackage{lipsum}
\usepackage{graphicx}
\usepackage{amsmath, bm}
\usepackage{caption}
\usepackage{multirow}

\theoremstyle{plain}

\theoremstyle{definition}

\theoremstyle{remark}

\usepackage[textsize=tiny]{todonotes}

\icmltitlerunning{Bayesian approaches for Quantifying Clinicians’ Variability in Medical Image Quantification}

\begin{document}

\twocolumn[
\icmltitle{Bayesian approaches for Quantifying Clinicians’ Variability in Medical Image Quantification}



\icmlsetsymbol{equal}{*}

\begin{icmlauthorlist}
\icmlauthor{Jaeik Jeon}{equal,cai}
\icmlauthor{Yeonggul Jang}{equal,cai,BK21}
\icmlauthor{Youngtaek Hong}{cai}
\icmlauthor{Hackjoon Shim}{cai}
\icmlauthor{Sekeun Kim}{hvd}
\end{icmlauthorlist}

\icmlaffiliation{cai}{CONNECT-AI Research Center, Yonsei University College
of Medicine, Seoul, South Korea}
\icmlaffiliation{BK21}{Graduate School of Medical Science, Brain Korea 21 Project, Yonsei University College of Medicine, Seoul, 03722, South Korea}
\icmlaffiliation{hvd}{Gordon Center for Medical Imaging, Massachusetts General
Hospital and Harvard Medical School, Boston, MA 02114, USA}

\icmlcorrespondingauthor{Sekeun Kim}{skim207@mgh.harvard.edu}

\icmlkeywords{Machine Learning, ICML}

\vskip 0.3in
]



\printAffiliationsAndNotice{\icmlEqualContribution} 

\begin{abstract}
  Medical imaging, including MRI, CT, and Ultrasound, plays a vital role in clinical decisions. Accurate segmentation is essential to measure the structure of interest from the image. However, manual segmentation is highly operator-dependent, which leads to high inter and intra- variability of quantitative measurements. In this paper, we explore the feasibility that Bayesian predictive distribution parameterized by deep neural networks can capture the clinicians' inter-intra variability. By exploring and analyzing recently emerged approximate inference schemes, we evaluate whether approximate Bayesian deep learning with the posterior over segmentations can learn inter-intra rater variability both in segmentation and clinical measurements. The experiments are performed with two different imaging modalities: MRI and ultrasound. We empirically demonstrated that Bayesian predictive distribution parameterized by deep neural networks could approximate the clinicians' inter-intra variability. We show a new perspective in analyzing medical images quantitatively by providing clinical measurement uncertainty.

\end{abstract}
\section{Introduction}
In quantitative medical image analysis, there is large inter-intra observer variability. Although this uncertainty arises from a variety of external sources, such as lack of precision in imaging devices or systematic measurement errors, manual segmentation is also dependent on the experience and perception of radiologists. The ambiguities of quantitative measurement has the potential to mislead poor clinical decision-making. Due to this uncertainty, there has been a clinician consensus that a deep learning model should provide distributions over possible outcomes rather than point estimates.

Most previous works on modeling uncertainty from inter/intra-rater variability focus on quantifying uncertainty at the per-pixel level. That is, many literatures aim to quantify clinicians' inter-intra variability in segmentation. \cite{baumgartner2019phiseg} proposed a probabilistic model that can generate segmentation samples closely resembling several annotators' ground-truth distribution. \citet{kendall2017uncertainties} proposed a method to quantify epistemic and aleatoric uncertainty in semantic segmentation tasks. In \citet{hu2019supervised}, they have defined aleatoric uncertainty to the per-pixel variance among the multiple segmentation by clinicians, built upon \citet{kohl2018probabilistic, gal2016dropout}. \citet{kohl2018probabilistic} proposed a probabilistic U-Net, a combination of a conditional variational auto-encoder and U-Net, for the segmentation of ambiguous images. It provides multiple plausible semantic segmentation hypotheses.


These prior works focus on returning calibrated uncertainty estimates to inform clinicians about the model confidence of its prediction. However, in quantitative medical image analysis, segmentation itself lacks clinical significance, and the uncertainty of segmentation is also lacking accordingly. Its endpoint is often to derive clinical measurements for disease diagnosis and decision-making, which is very important in the medical field. Therefore, in contrast to the previous approaches, we aim to model uncertainty on clinical indices, not per-pixel level uncertainty. Since the clinical indices such as volume and diameter are used in clinical decision in practice, the uncertainty of clinical indices has a greater clinical significance than pixel-wise uncertainty modeling. 
\begin{figure*}[t]
\centering
{
\includegraphics[width=1\textwidth]{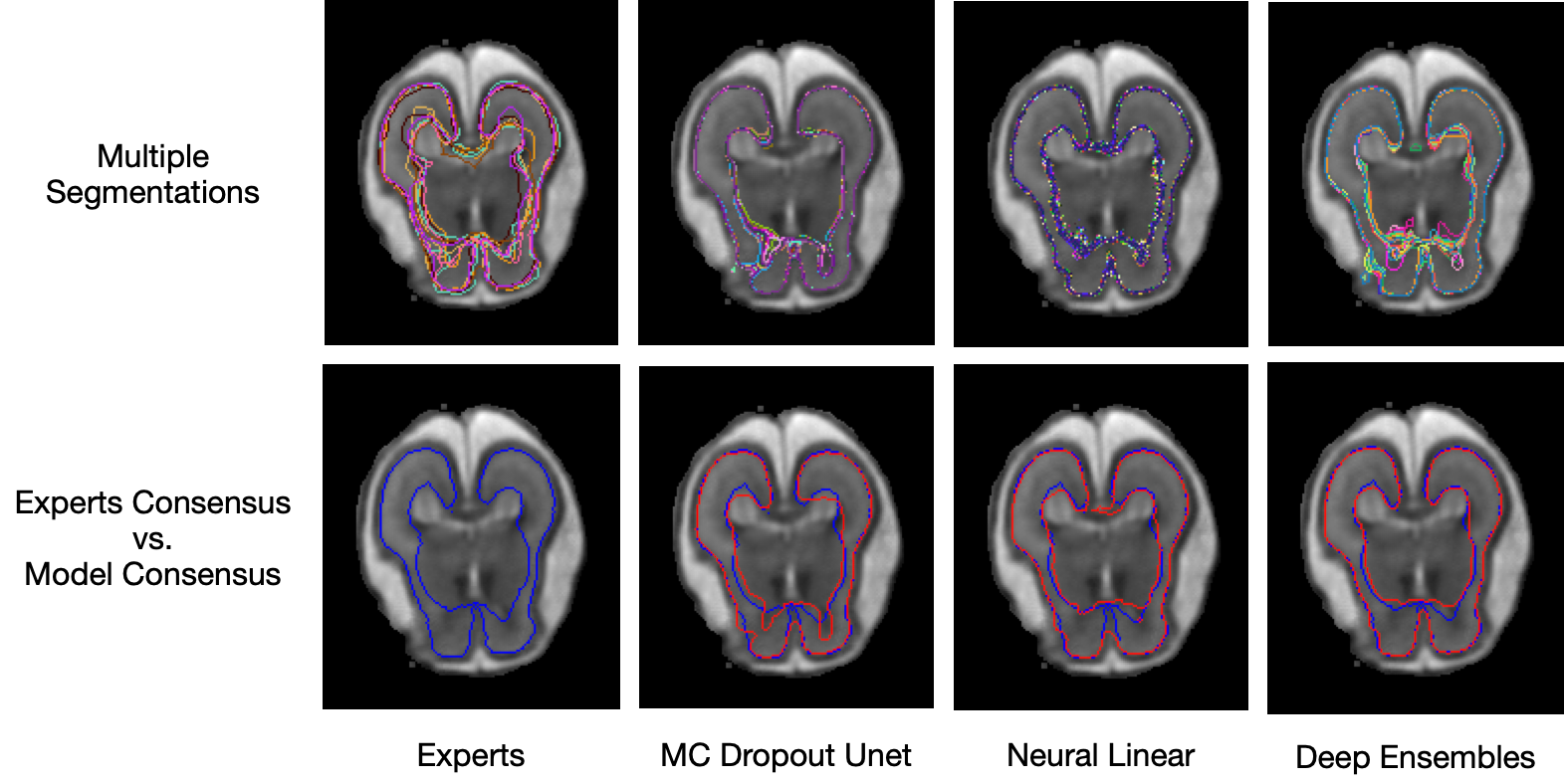}
}
\caption{Visualization of the learned variability in segmentation over the 2D MRI image. First rows presents multiple segmentations generated from experts and various Bayesian segmentation models. We see that all presented Bayesian networks successfully approximate the inter-variability of experts' annotations. Different annotations due to inter-rater variability are indicated by randomly selected colors.
} \label{fig1}
\end{figure*}

In this paper, we study the feasibility of a Bayesian predictive distribution parameterized by deep neural networks to model expert variability. By exploring and analyzing neural linear, MC dropout and deep ensembles, we evaluate whether approximate Bayesian deep learning with the posterior over segmentations can learn inter-intra rater variability both in segmentation and clinical measurements. The methods are evaluated on two datasets: a) the QUBIQ2021 dataset, which is used for evaluating the segmentation performance, and b) the IVUS dataset, which is used for evaluating a clinical endpoint. We do not propose new methods, but rather suggest that Bayesian approaches can approximate expert variability.

Our main contributions of this paper are:
\begin{enumerate}
    \item We explore and analyze three recently emerged Bayesian uncertainty estimation techniques that are scalable to the medical imaging segmentation task. 
    \item We show that the explored Bayesian methods are effective to quantify clinician's inter-intra variability in segmentation, but also in clinical measurements over the various medical image quantification tasks.
    \item We provide inclusion of clinical endpoints in the evaluation with different medical imaging modalities including MRI and Ultrasound using publicly available datasets.
\end{enumerate}


\section{Background and Related Works}

In this section, we describe a general Bayesian approach to model the variability of experts' annotation of medical images. We hypothesize that the predictive distribution learned from data matches the variability of the expert's annotations. Specifically, we exploit Bayesian neural networks that place a probability distribution over the weight parameters with the hope that the  distribution of predicted segmentation is identical to the distribution of multiple segmentation generated by multiple clinicians.

Let $\mathbf X=\{\mathbf x_i\}_{i=1}^N$ be the training inputs and $ Y=\{y_i\}_{i=1}^N$ be the outputs where $\mathbf x_i \in \mathbb R^d$ and $y_i \in \{0,1\}$. We express uncertainty on the model parameters $\bm\omega\in\Omega$ by defining a likelihood distribution $p(Y|\mathbf X, \bm\omega)$ and placing a prior distribution $p(\bm\omega)$. From Bayes' rule, we then achieve the posterior distribution over the parameter space
$    p(\bm\omega|\mathbf X, Y) = \frac{p(Y|\mathbf X, \bm\omega)p(\bm\omega)}{
    \int_\Omega p(Y|\mathbf X, \bm\omega)p(\bm\omega) d\bm\omega} $
and the predictive distribution for a new input $\mathbf x^*$
$p(y^*|\mathbf x^*, \mathbf X, Y) = \int_{\Omega} p(y^*|\mathbf x^*, \bm\omega)p(\bm\omega|\mathbf X, Y)d\bm\omega$. The tricky part is from the integration in the normalizing factor for which the closed form is often intractable, so is the predictive distribution.

In this work, we focus on a variational inference which has been extensively used in Bayesian deep learning community \cite{blundell2015weight, gal2016dropout, kingma2013auto} which recasts marginalisation (integration) as an optimisation problem. The posterior distribution is approximated by a variational distribution $q_{\bm\theta}(\bm\omega)$ parametrised by $\bm\theta$ such that the variational distribution is the closest distirbution to the posterior. We consider Kullback-Leibler (KL) divergence \cite{kullback1951information} $\textrm{KL}[q_{\bm\theta}(\bm\omega)||p(\bm\omega|\mathbf X, Y)]$ between $q_{\bm\theta}(\bm\omega)$ and $p(\bm\omega|\mathbf X, Y)$
as a measure of closeness. The optimal variational distribution $q_{\bm\theta}(\bm\omega)$ is achieved by minimizing KL divergence w.r.t the variational parameters $\bm\theta$, which is equivalent to maximizing evidence lower bound (ELBO) $\mathcal L(\bm\theta) = \mathbb E_{q_{\bm\theta}(\bm\omega)}[\log p(Y|\mathbf X, \bm\omega)] - \textrm{KL}[q_{\bm\theta}(\bm\omega)||p(\bm\omega)] $. In order to achieve practical gradient estimator of the ELBO, we reparametrize the random parameter $\bm\omega\sim q_{\bm\theta}(\bm\omega)$ as $\bm\omega=g(\bm\theta,\bm\epsilon),$ for a differentiable function $g(\cdot)$ and $\bm\epsilon$ being an auxiliary random variable drawn from $p(\bm\epsilon)$, introduced in \cite{kingma2013auto}.
By applying the reparametrization trick, we are able to get a differentiable MC estimator of the ELBO w.r.t.  $\bm\theta$ that is compatible with backpropagation:
\begin{equation}
    \hat{\mathcal L}(\bm\theta) =
    \frac1T \sum_{t=1}^T\log p(Y|\mathbf X, \bm\omega_t = g(\bm\theta, \bm\epsilon_t)) - \textrm{KL}[q_{\bm\theta}(\bm\omega)|| p(\bm\omega)]
    \label{MC_elbo}
\end{equation}

\section{Methods}
In this section, we will describe how we construct deep segmentation networks in Bayesian ways. We applied four scalable Bayesian methods created for different purposes in previous papers to Unet \cite{ronneberger2015u}.

\begin{minipage}{\columnwidth}
\begin{center}
\includegraphics[width=1\columnwidth]{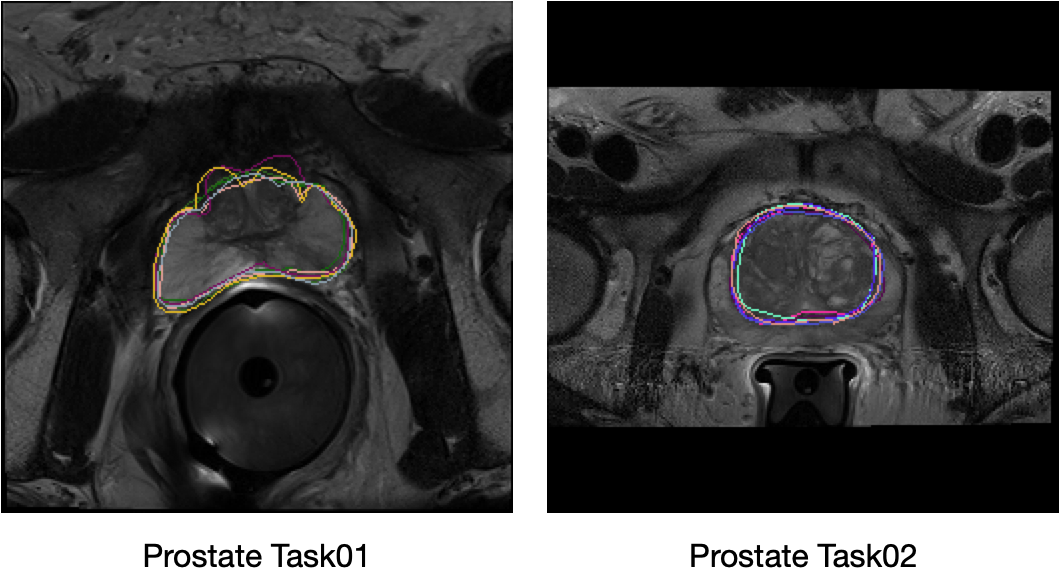}
\captionof{figure}{Visualization of the inter-variability over the prostate images in QUBIQ 2021 challenge.}
\end{center}
\end{minipage}

\subsection{Variational Inference}
\subsubsection{Estimate Uncertainty in the last layer}
\label{Estimate uncertainty in the last layer}
\citet{riquelme2018deep} performs Bayesian linear regression on features extracted from the last layer of the neural network to achieve accurate uncertainty estimates for Thompson sampling, which is named neural linear. 
 Inspired by the neural linear \cite{riquelme2018deep}, we employ Unet \cite{ronneberger2015u} as a deterministic feature extractor and  obtain  uncertainty estimates in the last layer. The last layer in Unet is a $1\times1$ convolution layer which projects the information over each channel to a class score, followed by the softmax function. We can think of this as a fully connected layer replicated for each pixel, which takes the corresponding channel vector as an input. By placing a probability distribution over the weights in the filter of the last layer, we capture the distribution in the aggregation process of information containing over channels in every pixel of the image.

We approximate the true posterior of weights in each filter with a fully factorised Gaussian invoking a mean field variational inference as in \citet{hinton1993keeping}:
\begin{equation}
    q_{\bm\theta}(\bm\omega) = \prod_{i=1}^I\prod_{j=1}^Oq_{\mu_{ij}\sigma_{ij}}(\omega_{ij})
    =\prod_{ij} N(\omega_{ij};\mu_{ij}, \sigma^2_{ij}),
\end{equation}
where $I$ is the number of input channel and $O$ number of output channel (classes). Following \cite{blundell2015weight}, we reparameterize $\bm\omega$ by $\omega_{ij} = \mu_{ij}+\sigma_{ij}\epsilon_{ij}$ with $\epsilon_{ij}\sim N(0,1)$. The standard deviation $\bm\sigma$ is further reparameterized by $\rho$ through a softplus function $\sigma_{ij} = \log(1+\exp(\rho_{ij}))$ to ensure $\sigma$ is always non-negative.

The local reparameterization trick \cite{kingma2015variational} is utilized for statistically efficient gradient estimation. Instead of sampling $\bm\omega$, we directly sample the random layer activation. Then the training procedure is described as follows:
\begin{enumerate}
    \item Sample $\epsilon_{ij}\sim N(0,1)$.
    \item Calculate $y_j=\sum_{i=1}^I z^{(lk)}_i\mu_{ij} + \epsilon_{ij} \sqrt{z^{{(lk)}^2}_i\sigma_{ij}} $, where $z^{(lk)}_i$ is the output from the unet feature extractor corresponding to $(lk)^{th}$ pixel.
    \item Calculate MC estimates of the ELBO according to (\ref{MC_elbo}).
    \item The deterministic neural network parameters and the variational parameters $\bm\theta=(\bm\mu, \bm\sigma)$ are updated through gradient ascent.
\end{enumerate}
When the training is done, the predictive distribution is achieved by MC approximation as follows:
\begin{equation}
    \hat p(y^*|\mathbf x^*, \mathbf X, Y) \approx \frac1T
    \sum_{t=1}^Tp(y^*|\mathbf x^*, \bm\omega_t),\:\: \bm\omega_t \sim q_{\bm\theta}(\bm\omega)
    \label{predictive distribtion}
\end{equation}


\subsubsection{MC dropout}
\citet{gal2016dropout} interprets the dropout training \cite{hinton2012improving,JMLR:v15:srivastava14a} as a variational inference which approximates the posterior distribution by placing Bernoulli variational distribution in the Bayesian neural networks. Dropout training is a commonly used regularization technique that prevents neural networks from overfitting and co-adaption of features in practice. It is done by randomly removing units within the neural networks during training. \citet{gal2016dropout} relates dropout with variational inference by defining the variational distribution $q_{\bm\theta}(\bm\omega)$ as a Bernoulli distribution for each layer in the network, and shows that the obtained model is the approximation of Gaussian processes \cite{gal2015bayesian}. The KL divergence between $q_{\bm\theta}(\bm\omega)$ and $p(\bm\omega)$ $\textrm{KL}[q_{\bm\theta}(\bm\omega)|| p(\bm\omega)]$ is minimized using stochastic gradient descent when train the network with cross entropy loss \cite{gal2015bayesian}. At test time, model parameters are sampled from the trained posterior distribution using dropout and the predictive distribution is achieved by the MC approximation as in equation \ref{predictive distribtion}.


\subsection{Deep Ensembles \cite{lakshminarayanan2016simple}}
In theory, Bayesian neural networks can capture uncertainty by learning a posterior distribution over the parameters of the network and exploring the space of solutions.
However, it has been reported that they often fail to explore the entire weight space and capture the network uncertainty within a single-mode \cite{fort2019deep}. Especially, \citet{fort2019deep} empirically shows that subspace sampling methods along with a single training trajectory exhibit high functional similarity, and the disagreement of predictions from the sampled functions is low.

Instead, deep neural networks initialized at a different random point tend to end up at different modes in function space \cite{fort2019deep, lakshminarayanan2016simple}. \citet{fort2019deep} shows that an ensemble of them that have different local minimums provides relatively larger benefits than subspace sampling methods such as MC dropout in terms of accuracy uncertainty in their experiment setting. We, therefore, construct an ensemble network that compensates for the insufficiency of a diverse set of predictions of the scalable variational inference methods by training the deterministic segmentation network multiple times with random initialization. The predictive distribution of an ensemble network then becomes
\begin{equation}
    \hat p(y^*|\mathbf x^*, \mathbf X, Y) \approx \frac1J
    \sum_{j=1}^Jp(y^*|\mathbf x^*,\hat{\bm\omega_j})
    \label{equation 4}
\end{equation}
where $\hat{\bm\omega_j}, \:\: j=1,\dots,J$ are $J$ independent model parameters trained from random initialization with different random seeds.

\section{Experiments and Results}

\begin{figure*}[t]
\centering
\includegraphics[width=0.7\textwidth]{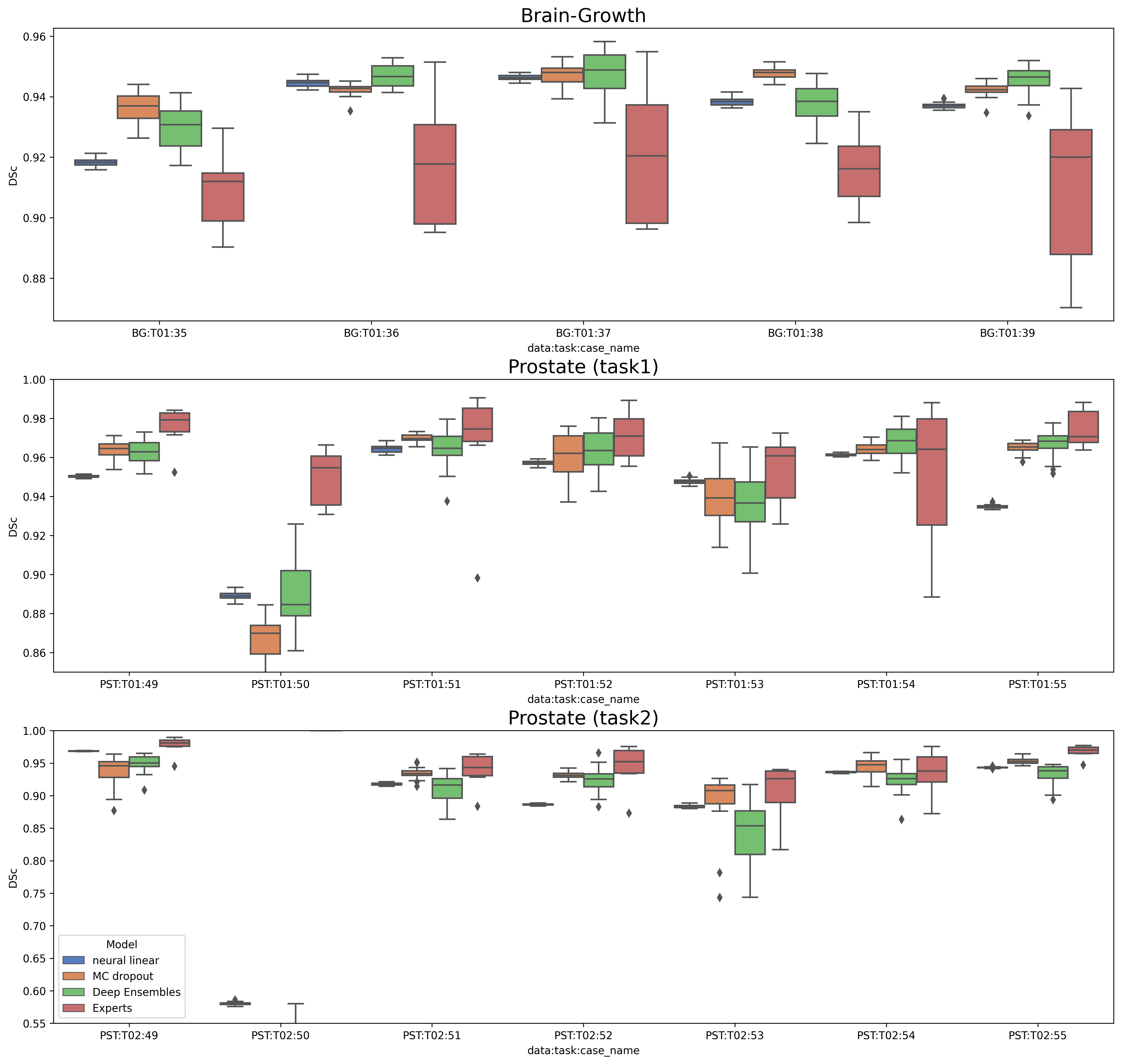}

\caption{Corresponding dice coefficient score between clinicians' consensus and sampled segmentation, but also the consensus and individual clinicians' annotation.}
 \label{fig3}
\end{figure*}
\begin{table*}[!t]
\centering
\caption{Generalized energy distance metric $D^2$ evaluations on MRI dataset.}
\resizebox{\textwidth}{!}{\begin{tabular}{c|c|c|c|cc}

Data sampling Strategy          & \multicolumn{1}{c|}{Methods}    & \multicolumn{1}{c|}{Prostate (task1)} & \multicolumn{1}{c|}{Prostate (task2)} & \multicolumn{1}{c|}{Brain-Growth (task1)} & average \\ \hline \centering
\multirow{1}{*}{All experts} & \multicolumn{1}{c|}{MC dropout \cite{kendall2015bayesian}} & \multicolumn{1}{c|}{0.117 $\pm$0.07}                 & \multicolumn{1}{c|}{0.305$\pm$ 0.44}                     & \multicolumn{1}{c|}{0.145 $\pm$0.012}                    &    0.194 $\pm$ 0.287     \\ \cline{2-6} 
                                & \multicolumn{1}{c|}{Neural Linear \cite{riquelme2018deep}}           & \multicolumn{1}{c|}{0.134 $\pm$ 0.05}                 & \multicolumn{1}{c|}{0.313 $\pm$ 0.33}                     & \multicolumn{1}{c|}{0.114 $\pm$ 0.02}                    &    0.194 $\pm$ 0.23     \\ \cline{2-6} 
                                &  Deep Ensembles \cite{fort2019deep}                               &   \bf{0.079 $\pm$0.05}                                    &         \bf{0.245 $\pm$ 0.36}                                  & \multicolumn{1}{c|}{\bf{0.104$\pm$0.01}}                    &     \bf{0.147 $\pm$ 0.24}    \\ \hline
\multirow{1}{*}{per expert model}     & MC dropout \cite{kendall2015bayesian}                      & 0.122 $\pm$ 0.07                                      &        0.34 $\pm 0.42$                                   & \multicolumn{1}{c|}{0.203 $\pm$ 0.04}                    & 0.223 $\pm$ 0.27        \\ \cline{2-6} 
                                &   Neural Linear   \cite{riquelme2018deep}                            &        0.187 $\pm$ 0.09                               &        0.398 $\pm$ 0.323                                   &  \multicolumn{1}{c|}{0.196 $\pm$ 0.05}                    &   0.264 $\pm$ 0.224      \\ \cline{2-6} 
                                & Deep Ensembles \cite{fort2019deep}                                &       \bf{    0.1 $\pm$ 0.05}                         &        \bf{     0.271 $\pm$ 0.32                }             & \multicolumn{1}{c|}{\bf{0.129 $\pm$ 0.02}    }                    &  \bf{      0.169 $\pm$ 0.2}
\end{tabular} \label{table: ged}}
\end{table*}

\subsection{Dataset} We use three datasets where images have multiple annotations from multiple clinicians. From QUBIQ 2021 challenge, the Prostate and Brain-Growth datasets were used, which have 7 and 6 inter-observer annotations respectively. The prostate dataset has two tasks that segment different regions of interest. We use single 2D MRI images size of 256 $\times$ 256 pixels. All pixels are normalized to [0, 1].
The annotated mask has binary labels: region of interest is 1, and background is 0. Sample brain-growth images with annotated masks are presented in figure \ref{fig1}. Inter-observer annotations are exists both in the train and test sets. The detailed training strategy is discussed in section \ref{sec: data sampling strategy}. Brain-growth dataset consists of 34 training set and 5 validation set, and prostate datset consists of 48 training set and 7 validation set.
We also use intravascular ultrasound (IVUS) datasets to demonstrate the applicability of predicting clinical indicators with ranges. The dataset consists of two sub-datasets, datasets A and B, which are publicly available \cite{balocco2014standardized}. Dataset A consists of 77 images with 19 for training and 58 for testing with a 40 MHz catheter. Dataset B consists of 425 images, with 109 for training and 326 for testing with a 20 MHz catheter. Unlike QUBIQ 2021 challenge dataset, the IVUS datset consists of 3 inter-intra observer annotations only in the test set. 
\begin{minipage}{\columnwidth}
\begin{center}
\includegraphics[width=1\columnwidth]{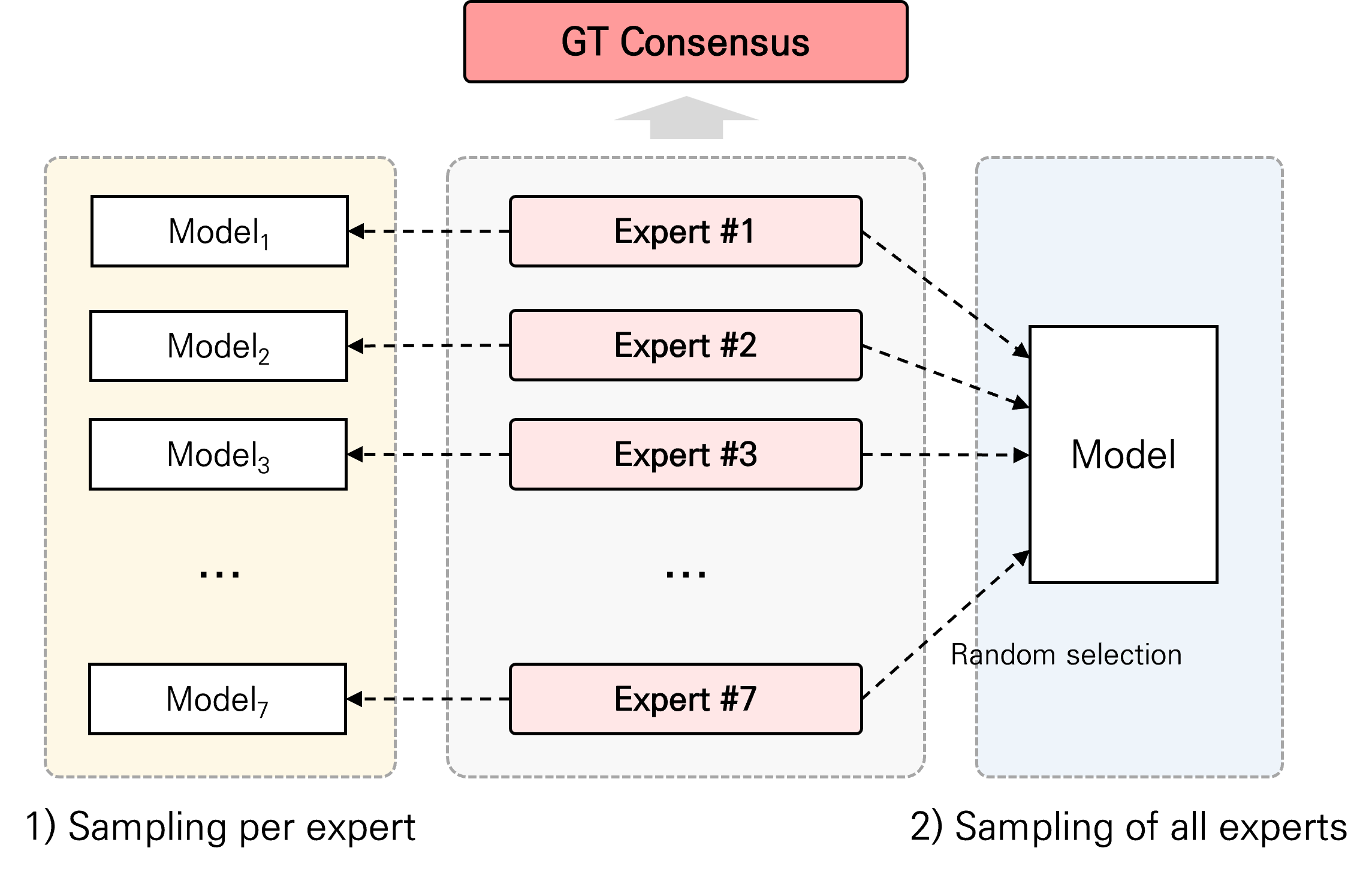}
\captionof{figure}{A schematic diagram of model training procedures using multiple annotated datasets from multiple experts.}
\end{center}
\label{figure:data_sampling_stategy}
\end{minipage}
\subsection{Implementation Details}
As described in \ref{Estimate uncertainty in the last layer}, motivated by \citet{riquelme2018deep}, we construct a neural linear segmentation network, performing Bayesian inference only in a small part of the network. The neural linear Unet comprises two main components: feature extractor and Bayesian convolution layer. From the Unet feature extractor, 64 feature maps are produced from each pixel of an image. The produced feature maps are fed into the stochastic $1\times1$ convolution layer. The Bayesian layer will capture the uncertainty by aggregating the information over 64 feature maps. The mean-field approximation is used for the variational posterior $q_{\bm\theta}(\bm\omega) = N(\omega_{ij};\mu_{ij}, \sigma_{ij}=\log(1+\exp(\rho_{ij})))$, where the variational parameters are initialized by $\mu_{ij}\sim N(0, 0.05)$,  $\rho_{ij}\sim N(-4, 0.05)$. This initialization is from \cite{pinsler2019bayesian}. We construct MC dropout Unet by allocating the dropout layer in the same Unet layer location as the \citet{kendall2015bayesian}. The only difference is that \citet{kendall2015bayesian} doesn't have a  skipconnection between encoder and decoder. We use 0.5 dropout rate both in training and test phase. For training, we use Adam optimizer \cite{kingma2014adam} with a learning rate 0.001 and cosine annealing is applied. The network is trained for 3000 epochs with mini-batch size 8 and early stopped at the best validation dice similarity coefficient (DSC).  No data augmentation is used. All architectures share consistent choices of other hyper-parameters that we explored. 20 MC samples and ensemble members are used throughout the experiments.

\subsection{Data Sampling Strategy}
\label{sec: data sampling strategy}

In this paper, we adopted two strategies for training models using datasets independently annotated by multiple experts: 1) Sampling per expert; 2) Sampling of all experts. First, we build the same number of models as experts, and each model learns only the annotations of one of the multiple experts. After all, this is the same as building each expert-specific model. Second, only one model is built, and the model learns one randomly selected from multiple annotations on the same data during training. Additionally, we combine multiple annotations to make an annotation consensus. Figure \ref{figure:data_sampling_stategy} shows a visual summary of these processes.

\subsection{Results Analysis}
Our analysis focuses on two aspects.
First, we validate the methods for approximating clinicians' variability by evaluating the similarity of clinicians' inter annotations and sampled masks from the predictive distribution using the QUBIQ2021 challenege dataset. In specific, we assess how similar the variances of the calculated  dice coefficients are from the learned predictive distributions to the clinicians' inter variability. 

Second, we evaluate a clinical endpoint of the learned posterior (uncertainty) over the segmentation using the IVUS dataset \cite{balocco2014standardized}.  We assess whether Bayesian approaches can approximate expert variability in the clinical measurement space. For a clinical endpoint, Lumen, EEM and plaque burden are calculated from the IVUS segmentation. We analyze whether the variance (clinical uncertainty) of the clinical values obtained through the Bayesian approaches and experts is consistent.

\subsubsection{Uncertainty Evaluation Metrics}

\begin{figure*}[t]
\centering
\includegraphics[width=0.6\textwidth]{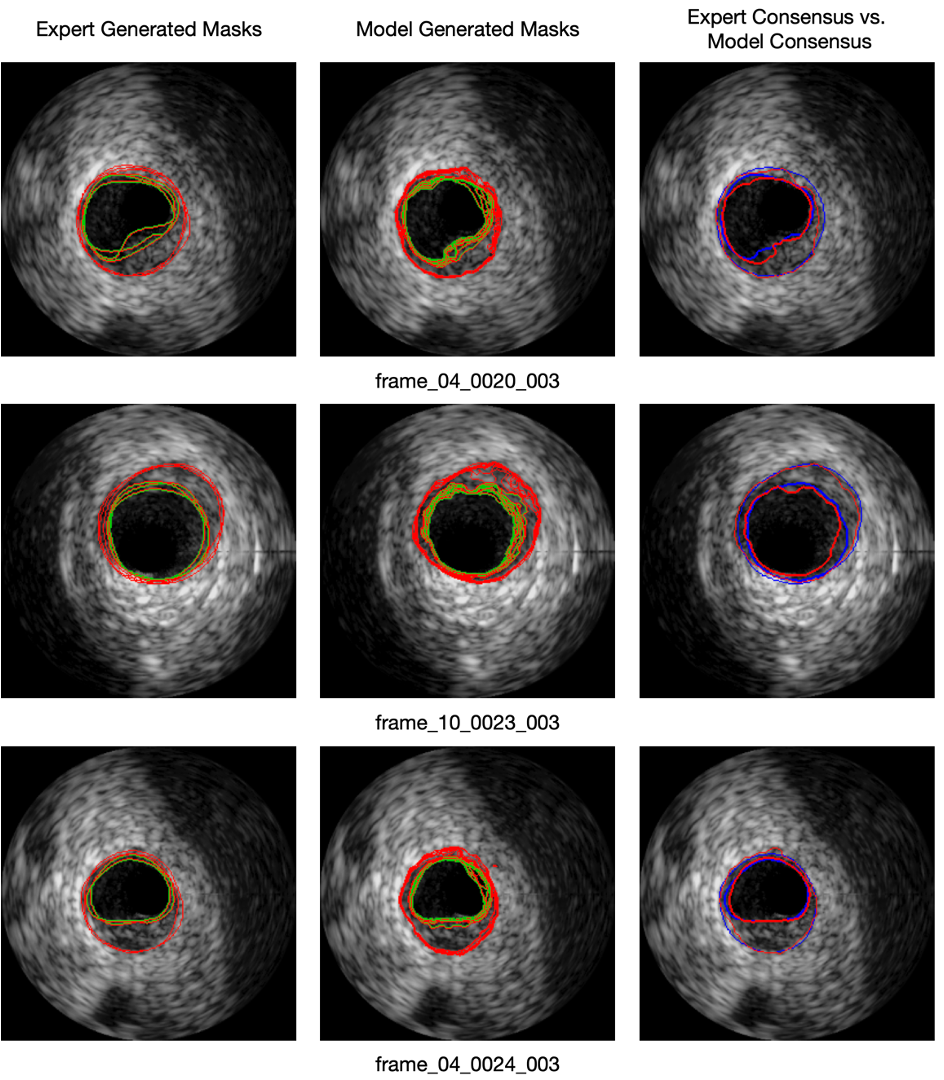}

\captionof{figure}{Visualization of the learned variability in segmentation over the IVUS dataset  \label{figure4} as well as inter-intra rater variability and their consensus.}
\captionof{table}{The uncertainty of (unscaled) clinical measurements for a single IVUS image in both experts and Bayesian models. \label{table5}}
\centering
\resizebox{\textwidth}{!}{
\begin{tabular}{llllllllll}
\toprule
& \multicolumn{3}{c}{patient 04 (frame \#20)} 
& \multicolumn{3}{c@{}}{patient 10 (frame \#23)}
& \multicolumn{3}{c@{}}{patient 04 (frame \#24)} \\
\cmidrule(lr){2-4} \cmidrule(l){5-7} \cmidrule(l){8-10}
& {Lumen} & {EEM} & {Plaque burden} & {Lumen} & {EEM} & {Plaque burden} & {Lumen} & {EEM} & {Plaque burden} \\
\midrule
Experts & 4081 $\pm$ 510 & 7138 $\pm 224$ & 0.43 $\pm$ 0.06 &5669 $\pm 202$ & 8675 $\pm 302$ & 0.346 $\pm$ 0.01 & 3190 $\pm 205$ & 5829 $\pm 153$ & 0.45 $\pm$0.02    \\
Deep Ensembles  & 4034 $\pm 232$ &  6212 $\pm 230$  & 0.39 $\pm$0.04 & 4772 $\pm$338 & 7671 $\pm$ 366 & 0.40 $\pm$0.04 & 2923 $\pm 166$ & 5370 $\pm 301$ &  0.47 $\pm 0.03$ \\
\bottomrule
\end{tabular}}
\end{figure*}

\textbf{Generalized Energy Distance Metric}
We aim to analyze if the approximate Bayesian deep learning with variational inference and deep ensembles can learn inter-rater variability in segmentation. To quantitatively assess the reproducibility of the rater variability of the method both in accuracy and diversity, we exploit the generalized energy distance metric \cite{kohl2018probabilistic, hu2019supervised} $D^2(p_{\textrm{gt}}, p_{\textrm{out}})=2\mathbb E[d(S,Y)] - \mathbb E[d(S,S')] - \mathbb E[d(Y,Y')]$ where $d$ is an intersection over union (IoU),  $Y$ and $Y'$ are ground truth sampled from clinicians, and $S$ and $S'$ are samples from the predictive distribution $\hat p$.
Note that the first term quantify accuracy and the rest of the term quantify diversity.


The average generalized energy distance $D^2$ of the presented methods (i.e., neural linear \cite{riquelme2018deep}, MC dropout \cite{kendall2015bayesian} and Deep ensembles \cite{fort2019deep} Unet) are described in table \ref{table: ged}. Noticeably, in all tasks, deep ensemble had an overwhelmingly lower GED score and better performance than all other methods such as MC dropout and neural linear. The table \ref{table: ged} represents that tendency of deep ensembles to explore diverse modes in function space leads to diversity also in the segmentation space.  Neither the neural linear nor the GED scores of mc dropout were particularly high or low for all three tasks. 
This result agrees with the argument of \citet{fort2019deep} in that the different modes of the function space provide a greater advantage over the subspace sampling method. Although this is consistent regardless of how the data is sampled from inter-rater annotations, it can be seen that the general performance of all experts model is higher than that of per expert model. This may be because the random selection of various inter-rater annotations for each image acts as a regularizer from overfitting.


\subsubsection{Does methods learn experts' variability?}

The presented method is capable of providing multiple segmentation results. We compared the presented models' variation with the posterior over the segmentation on three different medical datasets. In figure \ref{fig3} we plot the corresponding DSc score between clinicians' consensus and sampled segmentation, but also individual clinician's annotation and the consensus.

From figure \ref{fig3} it can be seen that the standard deviation of DSc between inter-raters is quite large, ranging from 0.01 to 0.04. This can be viewed as a meaningful estimate of the aleatoric uncertainty, the irreducible noise found in the data \cite{hu2019supervised}. The behavior of learned posteriors over the segmentation agrees with the intuition from \cite{fort2019deep}. Even though neural linear and MC dropout Unet obtain reasonable variability of Dice coefficient scores, they lack diversity. However, deep ensembles successfully approximate expert variability in all tasks. Not only is the variance of DSc between the clinician's agreement and the sample split close to the expert's variance, but the mean DSc score is also close enough to the expert's mean. This can also be confirmed in Figure \ref{fig1}. Figure \ref{fig1} is the visualization of the learned variability over the brain-growth image. Although MC dropout and neural linear approximate expert variability well enough, they are not as accurate and diverse as deep ensemble. From this, we conclude that deep ensembles perform best with respect to learning inter-rater variability in segmentation.


\subsubsection{Uncertainty in the Clinical Measurement Space}

\begin{figure*}[h]
\centering
\includegraphics[width=1\textwidth]{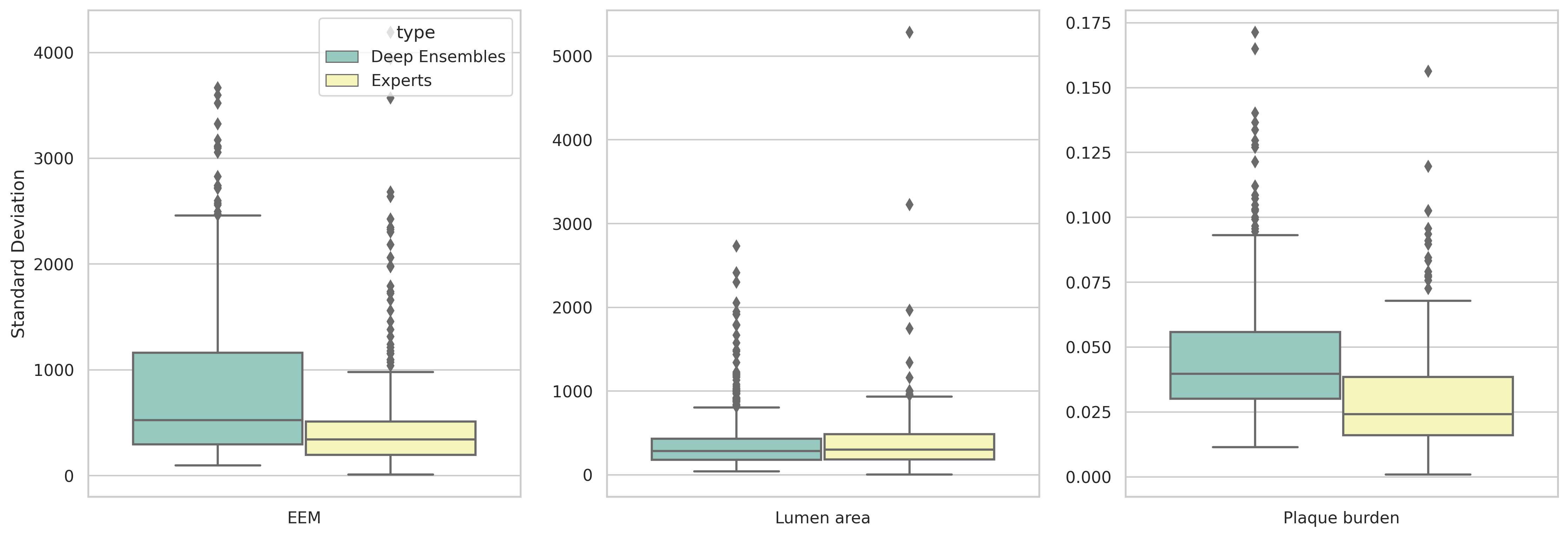}

\caption{Distributions of the standard deviation of the calculated measurements (EEM, lumen area, plaque burden) in a single IVUS image for all test dataset. \label{fig:figure7}}
\end{figure*}

\begin{minipage}{\columnwidth}
\begin{center}
\includegraphics[width=0.9\textwidth]{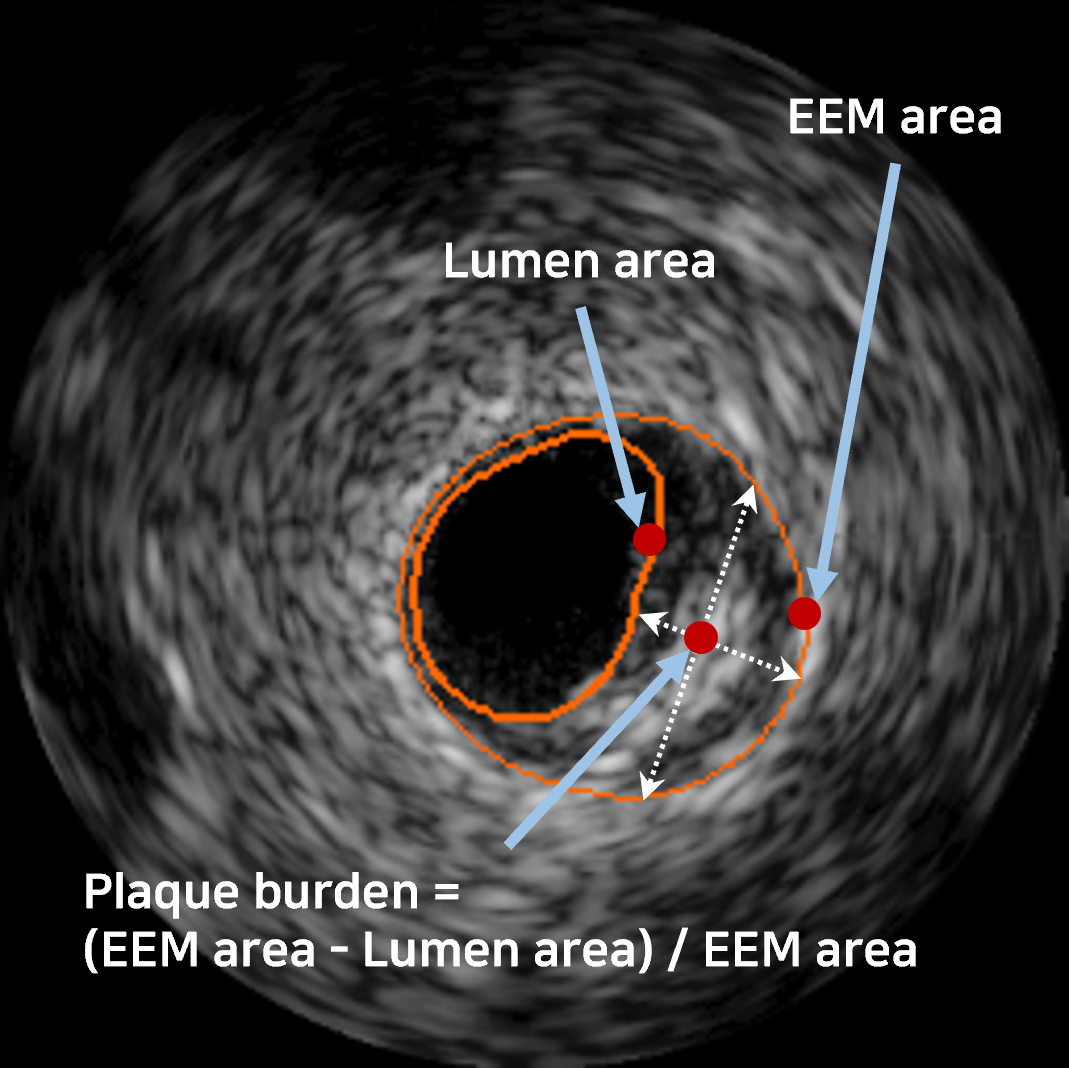}
\captionof{figure}{Visualization of measuring the cross-sectional plaque burden in IVUS imaging.}
\end{center}
\end{minipage}


In medical images, segmentation work does not end simply by itself. Its endpoint is to derive clinical measurements for disease diagnosis and decision-making, which is very important in the medical field. Therefore, to calculate the uncertainty of clinical measurements in the process, we propose to utilize multiple segmentations sampled from the posterior probability distribution, which is learned during training for the segmentation task.

In this section, we exploit the deep ensemble, best performed in previous experiments, to quantize uncertainty in clinical measurements and evaluate how effective it is. As the dataset for this experiment, we selected the IVUS dataset \cite{balocco2014standardized}, which is widely used as the gold standard for quantifying coronary artery stenosis. Coronary artery stenosis is quantified by measuring the cross-sectional plaque burden \cite{mcdaniel2011contemporary}. The plaque burden is calculated by dividing the plaque area by the external elastic membrane (EEM) area  as follows.

\begin{align}
\textrm{Plaque burden} = \frac{\left (\textrm{EEM} - \textrm{Lumen}\ \textrm{area} \right )}{\textrm{EEM}}\times 100
\end{align}

We first visualize the learned variability in segmentation over the IVUS dataset in figure \ref{figure4}.  As with the previous MRI dataset, learned posterior from deep ensembles successfully approximates the inter-variability in segmentation. In particular, from the patient 04 (frame \#20) in the first row, it can be seen that the variability of the depression of the lumen area is described as 
much as the inter-intra rater's variability. Furthermore, the variability of the boundary of the EEM area is also well approximated in all three cases. Figure \ref{table5} represents the uncertainty of (unscaled) clinical measurements for a single IVUS image in both experts and Bayesian models. Clinical indices such as EEM, lumen and plaque burden are described with error range by standard deviation. Although the estimate of the mean of each measurement is slightly different in some cases, an estimate of uncertainty by standard deviation is reasonable for comparison with an expert's standard deviation.  This information could be valuable tools in supporting clinicians' decision-making.

 
\begin{minipage}{\columnwidth}
\begin{center}
\includegraphics[width=0.8\textwidth]{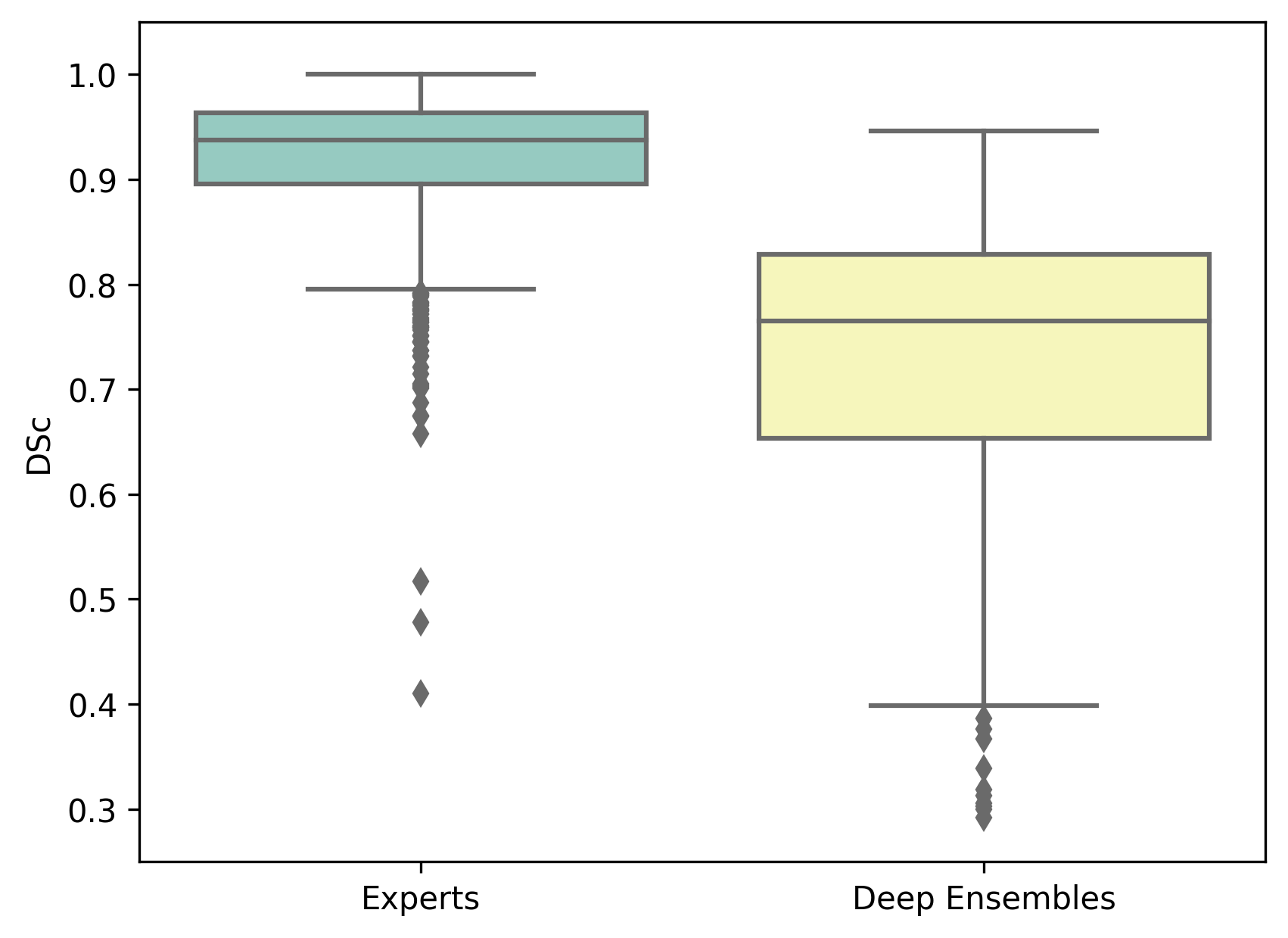}
\captionof{figure}{Distributions of dice coefficient between expert and expert consensus as well as trained model and expert consensus in IVUS test dataset. \label{figure9}}
\end{center}
\end{minipage}

We assess how well deep ensembles approximate the inter-intra variance (uncertainty) in the clinical measurements with figure \ref{fig:figure7}. It presents the distribution of the standard deviation of the calculated measurements (EEM, lumen area, plaque burden) in a single IVUS image for all test datasets. The mean and variance of the standard deviation of the predicted lumen areas within an image agree well with the experts' calculated values. The statistic for the EEM and plaque burden agree with the experts to some extent, but overall, the standard deviation is over-estimated. A clue to the overestimation can be found in the figure  \ref{figure9}, which shows distributions of dice coefficient between expert and expert consensus as well as the trained model and expert consensus. The DSc mean of experts is close to 0.93, whereas the DSc mean  of the deep ensembles is close to 0.79. This indicates that the model may be less trained for several reasons, leaving room for further performance gains with additional data. Therefore, we can hypothesize that the obtained predictive distributions lead to larger variances than the clinician's inter-intra measurements because of the combination of epistemic and aleatoric uncertainty. We leave it for future research to test this hypothesis and to analyze the results and performance of separating aleatoric and epistemic uncertainty for modeling the rater variability over the clinical measurements.


\section{Conclusion}
We analyze if Bayesian predictive distribution learned from various approximate inference schemes can learn intra-rater variability both in segmentation and clinical measurements. We do not propose new methods, but rather to demonstrate that Bayesian neural networks are able to reproduce the rater variability with the posterior over the segmentations in four medical imaging tasks. Especially, the IVUS dataset is used for evaluating a clinical endpoint. Our result suggests that Bayesian approaches can approximate expert variability.
For future research, we will analyze factors influencing uncertainty estimation performance in segmentation and clinical measurements. In specific, we will study how the epistemic and aleatoric uncertainties computed and separated by Bayesian neural networks in segmentation affect the approximation of uncertainty inherent in expert clinical measurements.






\bibliography{main}
\bibliographystyle{icml2022}



\end{document}